\documentclass[10pt]{article}
\usepackage[OE]{express}

\begin{document}

\title{Air mode silicon nitride photonic crystals and their application to nonlinear quantum optomechanical sensing}

\author{Chris Healey,\authormark{1,2} Hamidreza Kaviani,\authormark{1,2} and Paul E. Barclay\authormark{1,2,*}}

\address{\authormark{1}Institute For Quantum Science And Technology, University of Calgary, AB, Canada, T2N~1N4\\
\authormark{2}NRC National Institute For Nanotechnology, 11421 Saskatchewan Drive NW, Edmonton, AB, Canada, T6G~2M9}

\email{\authormark{*}pbarclay@ucalgary.ca} 



\begin{abstract}
Nanoscale photonic crystal cavity optomechanical devices enable detection of nanomechanical phenomena with a sensitivity sufficient to observe quantum effects.  Here we present the design of a one-dimensional air-mode photonic crystal cavity patterned in a silicon nitride nanobeam, and show that it forms the basis for cavity optomechanical split-beam and paddle nanocavity devices useful for force detection and nonlinear quantum sensing.  The air-mode of this device is advantageous for optomechanical coupling, while also having ultrahigh optical quality factor $Q_o\sim 10^6$ despite its proximity to the light-line and the relatively low refractive index of silicon nitride.  Paddle nanocavities realized from this device have a quadratic coupling coefficient $g^{(2)}/2\pi$~=~10~MHz/nm$^{2}$, and their performance within the context of quantum optomechanics experiments is analyzed.
\end{abstract}






\bibliographystyle{osajnl}
\bibliography{squidrefs}

\begin{thebibliography}{10}
\newcommand{\enquote}[1]{``#1''}

\bibitem{ref:aspelmeyer2014co}
M.~Aspelmeyer, T.~J. Kippenberg, and F.~Marquardt, \enquote{Cavity
  optomechanics,} Reviews of Modern Physics \textbf{86}, 1391--1452 (2014).

\bibitem{ref:favero2014fom}
I.~Favero and F.~Marquardt, \enquote{Focus on optomechanics,} New Journal of
  Physics \textbf{16}, 085006 (2014).

\bibitem{ref:metcalfe2014aco}
M.~Metcalfe, \enquote{Applications of cavity optomechanics,} Applied Physics
  Reviews \textbf{1}, 031105 (2014).

\bibitem{purdy2017quantum}
T.~Purdy, K.~Grutter, K.~Srinivasan, and J.~Taylor, \enquote{Quantum
  correlations from a room-temperature optomechanical cavity,} Science
  \textbf{356}, 1265--1268 (2017).

\bibitem{Chan2011}
J.~Chan, T.~P. Mayer~Alegre, A.~H. Safavi-Naeini, J.~T. Hill, A.~Krause,
  S.~Gr\"{o}blacher, M.~Aspelmeyer, and O.~Painter, \enquote{Laser cooling of a
  nanomechanical oscillator into its quantum ground state,} Nature
  \textbf{478}, 89--92 (2011).

\bibitem{Wu:17}
M.~Wu, N.~L.-Y. Wu, T.~Firdous, F.~{Fani Sani}, J.~E. Losby, M.~R. Freeman, and
  P.~E. Barclay, \enquote{Nanocavity optomechanical torque magnetometry and
  {RF} susceptometry,} Nature Nanotechnology \textbf{12}, 127 (2017).

\bibitem{Zhang:10}
Y.~Zhang, M.~Khan, Y.~Huang, J.~Ryou, P.~Deotare, R.~Dupuis, and M.~Lon\v{c}ar,
  \enquote{Photonic crystal nanobeam lasers,} Applied Physics Letters
  \textbf{97}, 051104 (2010).

\bibitem{Shakoor:14}
A.~Shakoor, K.~Nozaki, E.~Kuramochi, K.~Nishiguchi, A.~Shinya, and M.~Notomi,
  \enquote{Compact 1{D}-silicon photonic crystal electro-optic modulator
  operating with ultra-low switching voltage and energy,} Optics Express
  \textbf{22}, 28623--28634 (2014).

\bibitem{Chan:09}
J.~Chan, M.~Eichenfield, R.~Camacho, and O.~Painter, \enquote{Optical and
  mechanical design of a ``zipper'' photonic crystal optomechanical cavity,}
  Optics Express \textbf{17}, 3802--3817 (2009).

\bibitem{Quan10}
Q.~Quan, P.~Deotare, and M.~Lon\v{c}ar, \enquote{Photonic crystal nanobeam
  cavity strongly coupled to the feeding waveguide,} Applied Physics Letters
  \textbf{96}, 203102 (2010).

\bibitem{Quan11}
Q.~Quan and M.~Lon\v{c}ar, \enquote{Deterministic design of wavelength scale,
  ultra-high {Q} photonic crystal nanobeam cavities,} Optics Express
  \textbf{19}, 18529--18542 (2011).

\bibitem{Johnson02}
S.~G. Johnson, M.~Ibanescu, M.~A. Skorobogatiy, O.~Weisberg, J.~D.
  Joannopoulos, and Y.~Fink, \enquote{Perturbation theory for {Maxwell's}
  equations with shifting material boundaries,} Phys. Rev. E \textbf{65},
  066611 (2002).

\bibitem{Liang2015}
F.~Liang and Q.~Quan, \enquote{Detecting single gold nanoparticles (1.8 nm)
  with ultrahigh-{Q} air-mode photonic crystal nanobeam cavities,} ACS
  Photonics \textbf{2}, 1692--1697 (2015).

\bibitem{Yang:15}
D.~Yang, H.~Tian, and Y.~Ji, \enquote{High-{Q} and high-sensitivity
  width-modulated photonic crystal single nanobeam air-mode cavity for
  refractive index sensing,} Applied Optics \textbf{54}, 1--5 (2015).

\bibitem{Hryciw:13}
A.~C. Hryciw and P.~E. Barclay, \enquote{Optical design of split-beam photonic
  crystal nanocavities,} Optics Letters \textbf{38}, 1612--1614 (2013).

\bibitem{ref:wu2014ddo}
M.~Wu, A.~C. Hryciw, C.~Healey, D.~P. Lake, H.~Jayakumar, M.~R. Freeman, J.~P.
  Davis, and P.~E. Barclay, \enquote{Dissipative and dispersive optomechanics
  in a nanocavity torque sensor,} Physical Review~X \textbf{4}, 021052 (2014).

\bibitem{zhang2016chip}
X.~Zhang, G.~Zhou, P.~Shi, H.~Du, T.~Lin, J.~Teng, and F.~S. Chau,
  \enquote{On-chip integrated optofluidic complex refractive index sensing
  using silicon photonic crystal nanobeam cavities,} Optics Letters
  \textbf{41}, 1197--1200 (2016).

\bibitem{lin2015design}
T.~Lin, F.~Tian, P.~Shi, F.~S. Chau, G.~Zhou, X.~Tang, and J.~Deng,
  \enquote{Design of mechanically-tunable photonic crystal split-beam
  nanocavity,} Optics Letters \textbf{40}, 3504--3507 (2015).

\bibitem{Healey:15}
C.~Healey, H.~Kaviani, M.~Wu, B.~Khanaliloo, M.~Mitchell, A.~C. Hryciw, and
  P.~E. Barclay, \enquote{Design and experimental demonstration of
  optomechanical paddle nanocavities,} Applied Physics Letters \textbf{107},
  231107 (2015).

\bibitem{Kaviani:15}
H.~Kaviani, C.~Healey, M.~Wu, R.~Ghobadi, A.~Hryciw, and P.~E. Barclay,
  \enquote{Nonlinear optomechanical paddle nanocavities,} Optica \textbf{2},
  271--274 (2015).

\bibitem{Baets:16}
R.~G.~F. Baets, A.~Z. Subramanian, S.~Clemmen, B.~Kuyken, P.~Bienstman, N.~L.
  Thomas, G.~Roelkens, D.~V. Thourhout, P.~Helin, and S.~Severi,
  \enquote{Silicon photonics: Silicon nitride versus silicon-on-insulator,} in
  \enquote{Optical Fiber Communication Conference,}  (Optical Society of
  America, 2016), p. Th3J.1.

\bibitem{Verbridge:06}
S.~S. Verbridge, J.~M. Parpia, R.~B. Reichenbach, L.~M. Bellan, and H.~G.
  Craighead, \enquote{High quality factor resonance at room temperature with
  nanostrings under high tensile stress,} Journal of Applied Physics
  \textbf{99}, 124304 (2006).

\bibitem{Verbridge:08}
S.~S. Verbridge, H.~G. Craighead, and J.~M. Parpia, \enquote{A megahertz
  nanomechanical resonator with room temperature quality factor over a
  million,} Applied Physics Letters \textbf{92}, 013112 (2008).

\bibitem{ghadimi2018elastic}
A.~H. Ghadimi, S.~A. Fedorov, N.~J. Engelsen, M.~J. Bereyhi, R.~Schilling,
  D.~J. Wilson, and T.~J. Kippenberg, \enquote{Elastic strain engineering for
  ultralow mechanical dissipation,} Science \textbf{360}, 764--768 (2018).

\bibitem{Camacho:09}
R.~M. Camacho, J.~Chan, M.~Eichenfield, and O.~Painter,
  \enquote{Characterization of radiation pressure and thermal effects in a
  nanoscale optomechanical cavity,} Optics Express \textbf{17}, 15726--15735
  (2009).

\bibitem{Krause:12}
A.~G. Krause, M.~Winger, T.~D. Blasius, Q.~Lin, and O.~Painter, \enquote{A
  high-resolution microchip optomechanical accelerometer,} Nature Photonics
  \textbf{6}, 768--772 (2012).

\bibitem{Krause:15}
A.~Krause, T.~Blasius, and O.~Painter, \enquote{Optical read out and feedback
  cooling of a nanostring optomechanical cavity,} arXiv:1506.01249  (2015).

\bibitem{norte2016mechanical}
R.~A. Norte, J.~P. Moura, and S.~Gr{\"o}blacher, \enquote{Mechanical resonators
  for quantum optomechanics experiments at room temperature,} Physical Review
  Letters \textbf{116}, 147202 (2016).

\bibitem{reinhardt2016ultralow}
C.~Reinhardt, T.~M{\"u}ller, A.~Bourassa, and J.~C. Sankey,
  \enquote{Ultralow-noise {SiN} trampoline resonators for sensing and
  optomechanics,} Physical Review~X \textbf{6}, 021001 (2016).

\bibitem{tsaturyan2017ultracoherent}
Y.~Tsaturyan, A.~Barg, E.~S. Polzik, and A.~Schliesser, \enquote{Ultracoherent
  nanomechanical resonators via soft clamping and dissipation dilution,} Nature
  Nanotechnology  (2017).

\bibitem{chang2017heterogeneous}
L.~Chang, M.~H. Pfeiffer, N.~Volet, M.~Zervas, J.~D. Peters, C.~L. Manganelli,
  E.~J. Stanton, Y.~Li, T.~J. Kippenberg, and J.~E. Bowers,
  \enquote{Heterogeneous integration of lithium niobate and silicon nitride
  waveguides for wafer-scale photonic integrated circuits on silicon,} Optics
  Letters \textbf{42}, 803--806 (2017).

\bibitem{Barth:08}
M.~Barth, N.~N\"{u}sse, J.~Stingl, B.~L\"{o}chel, and O.~Benson,
  \enquote{Emission properties of high-{Q} silicon nitride photonic crystal
  heterostructure cavities,} Applied Physics Letters \textbf{93}, 021112
  (2008).

\bibitem{McCutcheon:08}
M.~W. McCutcheon and M.~Lon\v{c}ar, \enquote{Design of a silicon nitride
  photonic crystal nanocavity with a quality factor of one million for coupling
  to a diamond nanocrystal,} Optics Express \textbf{16}, 19136--19145 (2008).

\bibitem{panettieri2016control}
D.~Panettieri, L.~O'Faolain, and M.~Grande, \enquote{Control of {Q}-factor in
  nanobeam cavities on substrate,} in \enquote{2016 18th International
  Conference on Transparent Optical Networks~(ICTON),}  (IEEE, 2016), pp. 1--4.

\bibitem{fryett2018encapsulated}
T.~K. Fryett, Y.~Chen, J.~Whitehead, Z.~M. Peycke, X.~Xu, and A.~Majumdar,
  \enquote{Encapsulated silicon nitride nanobeam cavity for hybrid
  nanophotonics,} ACS {P}hotonics \textbf{5}, 2176--2181 (2018).

\bibitem{chen2018deterministic}
Y.~Chen, A.~Ryou, M.~R. Friedfeld, T.~Fryett, J.~Whitehead, B.~M. Cossairt, and
  A.~Majumdar, \enquote{Deterministic positioning of colloidal quantum dots on
  silicon nitride nanobeam cavities,} Nano Letters \textbf{18}, 6404--6410
  (2018).

\bibitem{PCbook}
J.~D. Joannopoulos, S.~G. Johnson, J.~N. Winn, and R.~D. Meade, \emph{Photonic
  Crystals: Molding the Flow of Light second edition} (Princeton University
  Press, 2008).

\bibitem{Khan:11}
M.~Khan, T.~Babinec, M.~W. McCutcheon, P.~Deotare, and M.~Lon\v{c}ar,
  \enquote{Fabrication and characterization of high-quality-factor silicon
  nitride nanobeam cavities,} Optics Letters \textbf{36}, 421--423 (2011).

\bibitem{Davanco:14}
M.~Davan\c{c}o, S.~Ates, Y.~Liu, and K.~Srinivasan, \enquote{Si$_3${N}$_4$
  optomechanical crystals in the resolved-sideband regime,} Applied Physics
  Letters \textbf{104}, 041101 (2014).

\bibitem{Barth:07}
M.~Barth, J.~Kouba, J.~Stingl, B.~L\"{o}chel, and O.~Benson,
  \enquote{Modification of visible spontaneous emission with silicon nitride
  photonic crystal nanocavities,} Optics Express \textbf{15}, 17231--17240
  (2007).

\bibitem{Eichenfield:09}
M.~Eichenfield, R.~Camacho, J.~Chan, K.~J. Vahala, and O.~Painter, \enquote{A
  picogram- and nanometre-scale photonic-crystal optomechanical cavity,} Nature
  \textbf{459}, 550--555 (2009).

\bibitem{Grutter:15}
K.~E. Grutter, M.~I. Davan\c{c}o, and K.~Srinivasan, \enquote{Slot-mode
  optomechanical crystals: a versatile platform for multimode optomechanics,}
  Optica \textbf{2}, 994--1001 (2015).

\bibitem{Mouradian:17}
S.~Mouradian, N.~H. Wan, T.~Schr\"{o}der, and D.~Englund, \enquote{Rectangular
  photonic crystal nanobeam cavities in bulk diamond,} Applied Physics Letters
  \textbf{111}, 021103 (2017).

\bibitem{Johnson:01}
S.~G. Johnson and J.~D. Joannopoulos, \enquote{Block-iterative frequency-domain
  methods for {M}axwell's equations in a planewave basis,} Optics Express
  \textbf{8}, 173--190 (2001).

\bibitem{ref:oskooi2010mff}
A.~F. Oskooi, D.~Roundy, M.~Ibanescu, P.~Bermel, J.~D. Joannopoulos, and S.~G.
  Johnson, \enquote{{MEEP}: {A} flexible free-software package for
  electromagnetic simulations by the {FDTD} method,} Computer Physics
  Communications \textbf{181}, 687--702 (2010).

\bibitem{Grutter:15IEEE}
K.~E. Grutter, M.~Davan\c{c}o, and K.~Srinivasan, \enquote{Si$_3${N}$_4$
  nanobeam optomechanical crystals,} IEEE Journal Of Selected Topics In Quantum
  Electronics \textbf{21}, 2700611--2700611 (2015).

\bibitem{Zhang:09}
Y.~Zhang, M.~W. McCutcheon, I.~B. Burgess, and M.~Lon\v{c}ar,
  \enquote{Ultra-high-{Q} {TE}/{TM} dual-polarized photonic crystal
  nanocavities,} Optics Letters \textbf{34}, 2\,694--2\,696 (2009).

\bibitem{goh2007genetic}
J.~Goh, I.~Fushman, D.~Englund, and J.~Vu{\v{c}}kovi{\'c}, \enquote{Genetic
  optimization of photonic bandgap structures,} Optics Express \textbf{15},
  8218--8230 (2007).

\bibitem{Thompson2008}
J.~D. Thompson, B.~M. Zwickl, A.~M. Jayich, F.~Marquardt, S.~M. Girvin, and
  J.~G.~E. Harris, \enquote{Strong dispersive coupling of a high-finesse cavity
  to a micromechanical membrane,} Nature \textbf{452}, 72--75 (2008).

\bibitem{sankey2010strong}
J.~C. Sankey, C.~Yang, B.~M. Zwickl, A.~M. Jayich, and J.~G. Harris,
  \enquote{Strong and tunable nonlinear optomechanical coupling in a low-loss
  system,} Nature Physics \textbf{6}, 707--712 (2010).

\bibitem{mason2018continuous}
D.~Mason, J.~Chen, M.~Rossi, Y.~Tsaturyan, and A.~Schliesser,
  \enquote{Continuous force and displacement measurement below the standard
  quantum limit,} arXiv:1809.10629  (2018).

\bibitem{rossi2018observing}
M.~Rossi, D.~Mason, J.~Chen, and A.~Schliesser, \enquote{Observing and
  verifying the quantum trajectory of a mechanical resonator,} arXiv:1812.00928
   (2018).

\bibitem{rossi2018measurement}
M.~Rossi, D.~Mason, J.~Chen, Y.~Tsaturyan, and A.~Schliesser,
  \enquote{Measurement-based quantum control of mechanical motion,} Nature
  \textbf{563}, 53 (2018).

\bibitem{paraiso2015position}
T.~K. Para{\"\i}so, M.~Kalaee, L.~Zang, H.~Pfeifer, F.~Marquardt, and
  O.~Painter, \enquote{Position-squared coupling in a tunable photonic crystal
  optomechanical cavity,} Physical Review~X \textbf{5}, 041024 (2015).

\bibitem{Rodriguez:11}
A.~W. Rodriguez, A.~P. McCauley, P.-C. Hui, D.~Woolf, E.~Iwase, F.~Capasso,
  M.~Lon\v{c}ar, and S.~G. Johnson, \enquote{Bonding, antibonding and tunable
  optical forces in asymmetric membranes,} Optics Express \textbf{19},
  2225--2241 (2011).

\bibitem{Eichenfield:09OE}
M.~Eichenfield, J.~Chan, A.~H. Safavi-Naeini, K.~J. Vahala, and O.~Painter,
  \enquote{Modeling dispersive coupling and losses of localized optical and
  mechanical modes in optomechanical crystals,} Optics Express \textbf{17},
  20078--20098 (2009).

\bibitem{Hryciw:15}
A.~C. Hryciw, M.~Wu, B.~Khanaliloo, and P.~E. Barclay, \enquote{Tuning of
  nanocavity optomechanical coupling using a near-field fiber probe,} Optica
  \textbf{2}, 491--496 (2015).

\bibitem{Cleland:02}
A.~N. Cleland and M.~L. Roukes, \enquote{Noise processes in nanomechanical
  resonators,} Journal of Applied Physics \textbf{92}, 2758--2769 (2002).

\bibitem{ClerkPRL2010}
A.~A. Clerk, F.~Marquardt, and J.~G.~E. Harris, \enquote{Quantum measurement of
  phonon shot noise,} Physical Review Letters \textbf{104}, 213603 (2010).

\bibitem{miao2009standard}
H.~Miao, S.~Danilishin, T.~Corbitt, and Y.~Chen, \enquote{Standard quantum
  limit for probing mechanical energy quantization,} Physical review letters
  \textbf{103}, 100402 (2009).

\bibitem{ludwig2012enhanced}
M.~Ludwig, A.~H. Safavi-Naeini, O.~Painter, and F.~Marquardt, \enquote{Enhanced
  quantum nonlinearities in a two-mode optomechanical system,} Physical Review
  Letters \textbf{109}, 063601 (2012).

\bibitem{sekoguchi2014photonic}
H.~Sekoguchi, Y.~Takahashi, T.~Asano, and S.~Noda, \enquote{Photonic crystal
  nanocavity with a q-factor of\~{} 9 million,} Optics Express \textbf{22},
  916--924 (2014).

\bibitem{Clark2017}
J.~B. Clark, F.~Lecocq, R.~W. Simmonds, J.~Aumentado, and J.~D. Teufel,
  \enquote{Sideband cooling beyond the quantum backaction limit with squeezed
  light,} Nature \textbf{452}, 191--195 (2017).

\bibitem{ghadimi2017radiation}
A.~H. Ghadimi, D.~J. Wilson, and T.~J. Kippenberg, \enquote{Radiation and
  internal loss engineering of high-stress silicon nitride nanobeams,} Nano
  Letters \textbf{17}, 3501--3505 (2017).

\bibitem{fedorov2018evidence}
S.~Fedorov, V.~Sudhir, R.~Schilling, H.~Sch{\"u}tz, D.~J. Wilson, and
  T.~Kippenberg, \enquote{Evidence for structural damping in a high-stress
  silicon nitride nanobeam and its implications for quantum optomechanics,}
  Physics Letters A \textbf{382}, 2251--2255 (2018).

\end{thebibliography}

\section{Introduction}

One-dimensional photonic crystal cavities (PCC's) patterned in nanobeams have many uses in both fundamental and practical applications of cavity optomechanics~\cite{ref:aspelmeyer2014co, ref:favero2014fom, ref:metcalfe2014aco}, for instance to observe quantum correlations at room temperature~\cite{purdy2017quantum}, cooling of a nanomechanical oscillator into its quantum ground state~\cite{Chan2011}, ultra-sensitive torque magnetometry~\cite{Wu:17}, and to realize laser~\cite{Zhang:10} and electro-optic modulator technologies~\cite{Shakoor:14}.  Many methodologies for designing defects within nanobeam waveguides typically patterned with holes to define PCCs have been discussed. One of the most successful approaches relies on tailoring the envelope of the electric field in the central region of the nanobeam by gradually varying the hole shape or spacing~\cite{Chan:09}. By following a deterministic design procedure based on gradually varying the bandgap of each unit cell of the nanobeam~\cite{Quan10,Quan11}, optical modes  with ultra-high quality factors ($Q_o$) can be created with fields concentrated in the  high-index dielectric or lower-index air/vacuum regions, referred to as dielectric- and air-modes, respectively.

Cavities supporting air-modes are advantageous for applications such as sensing and optomechanics due to their high concentration of optical energy in the lower index regions surrounding the mechanically moving boundaries of the device \cite{Johnson02}.  Air-mode PCC's have been demonstrated with $Q_o$~\textgreater~$10^5$ at optical~\cite{Liang2015} frequencies, and have applications such as temperature and refractive index sensing~\cite{Yang:15} and detecting single nanoparticles~\cite{Liang2015}.  They are also the basis for optomechanical devices such as split-beam cavities~(SBC's)~\cite{Hryciw:13, ref:wu2014ddo, zhang2016chip, lin2015design}, where a complete gap is introduced into the centre of the photonic crystal, and paddle nanocavities~(PNC's)~\cite{Healey:15,Kaviani:15}, where a nano-mechanical resonator is inserted into the gap of the split-beam cavity and interacts strongly with the optical mode.  In these devices, the electric fields concentrated in the air and overlapping with the boundaries of the nanomechanical resonator allow sensitive transduction of the mechanical motion and enable studies of magnetometry~\cite{Wu:17} and nonlinear optomechanics~\cite{Kaviani:15}.  

To date, one dimensional air-mode PCC's have typically been fabricated from high refractive index silicon (Si, $n_\text{Si}~\sim~3.5$).   There is significant interest however in creating air mode silicon nitride devices (Si$_{3}$N$_{4}$,~$n_\text{Si$_{3}$N$_{4}$}~\sim~2.0$)~\cite{Baets:16}.  The large internal tensile stress of Si$_{3}$N$_{4}$ films on Si ($\sim$~1~GPa) allows for realization of high mechanical quality factor $Q_m$~\textgreater~$10^6$ nanobeam mechanical resonators~\cite{Verbridge:06, Verbridge:08, ghadimi2018elastic}. It also enables creating optomechanical structures~\cite{Camacho:09, Krause:12, Krause:15, norte2016mechanical, reinhardt2016ultralow, tsaturyan2017ultracoherent, ghadimi2018elastic} suspended over large length scales without warping or stiction compared with materials such as Si. This allows fabrication of long, thin supports suitable for suspending nanomechanical elements with enhanced sensitivity to external forces owing to a combination of low mechanical frequency~$\omega_m$, effective mass~$m_{eff}$, and spring-constant \cite{reinhardt2016ultralow}. For example, zipper cavities consisting of pairs of side-coupled dielectric-mode Si$_{3}$N$_{4}$ PCC's~\cite{Chan:09} have been fabricated with high aspect ratio tethers (115~$\mu$m~x~130~nm) for ultrasensitive accelerometers~\cite{Krause:12} and studying feedback cooling~\cite{Krause:15}.  As well, Si$_{3}$N$_{4}$ is compatible with standard silicon and CMOS manufacturing processes, has low optical absorption from telecom through most visible wavelengths with nominally no two-photon absorption at telecommunication wavelengths, and can be integrated with lithium niobate~\cite{chang2017heterogeneous}.  However, its relatively low refractive index is detrimental to high-$Q_o$ air-mode PCC design~\cite{Barth:08, McCutcheon:08} due to its reduced optical energy confinement and resulting increased coupling to vertically radiating modes.

Air-mode PCC's locally confine the higher energy conduction band mode of a photonic crystal waveguide through tapering a decreasing fill factor towards the centre of the crystal, and vice-versa for the dielectric mode. 
Encapsulated air-mode Si$_{3}$N$_{4}$ PCC's with elliptical holes designed for 740-1000~nm wavelengths have been used to show the deterministic positioning of nanoparticles~\cite{panettieri2016control,fryett2018encapsulated,chen2018deterministic}.  However, most demonstrations of Si$_{3}$N$_{4}$ PCC's to date have utilized the dielectric mode, ensuring that the bandedge frequency of the photonic crystal waveguide from which the PCC is formed is far from the light-line $\omega~=~ck_{x}$, where $k_x$ is component of the wavevector~$k$ parallel to the PCC's waveguiding axis~\cite{PCbook}.  Lowering the bandedge frequency from which the PCC mode is formed, either by increasing the effective refractive index~$n_{eff}$ or basing a PCC on a dielectric mode, reduces coupling between the optical mode and radiation modes above the light-line, and thus reduces loss.  Dielectric mode Si$_{3}$N$_{4}$ PCC's have been shown to have high $Q_o$~\cite{Khan:11}, have been demonstrated in the resolved-sideband regime of cavity optomechanics~\cite{Davanco:14}, and have been used to study the emission properties of SiN~\cite{Barth:08} and to modify spontaneous emission~\cite{Barth:07}.  Dielectric mode zipper cavities have also been used for cavity optomechanics~\cite{Eichenfield:09}, and side-coupled optical and mechanical beams can form slot-mode optomechanical crystals that have been used in multimode optomechanics~\cite{Grutter:15}.  These devices are often designed at target wavelengths for specific applications, for instance at 637~nm for interfacing with diamond color centres~\cite{McCutcheon:08}.  

In contrast, free-standing air-mode Si$_{3}$N$_{4}$ PCC's at telecommunication wavelengths have not been reported to the best of the authors' knowledge.  Here we show that air-mode Si$_{3}$N$_{4}$ nanobeam PCC's based on tapered round holes can be designed with ultra-high $Q_o$.  This is achieved by increasing the thickness of the devices, thus increasing $n_{eff}$ and lowering the air-mode band-edge frequency away from the light-line.  Motivated by high~$Q_o$ optomechanical SBC's and PNC's~\cite{Hryciw:13,Healey:15}, we then design devices based on elliptical holes that mode-match to a physical gap introduced within the PCC.  Due to the higher optical bandedge frequencies of these devices, the standard deterministic design approach fails to predict the highest $Q_o$ device. However, by substituting a highly elliptical hole for the centre hole and tapering to circular holes we can lower the band frequencies and achieve an order of magnitude higher $Q_o$ than predicted with the deterministic procedure.  We then further extend this highly elliptical hole design to add mechanical degrees of freedom and create high-$Q_o$ SBC and PNC designs.  These devices combine  the excellent mechanical attributes of Si$_{3}$N$_{4}$ with the sensitive cavity-optomechanical transduction of earlier Si based designs \cite{Kaviani:15}. This is predicted to allow their room-temperature thermal motion to be detected in a nonlinear optomechanical measurement whose power spectral density~(PSD) is observable above system noise levels.  We also show that in cryogenic conditions, observations of thermally-driven nonlinear optomechanical phononic Fock state quantum jumps and shot noise should be detectable.  Notwithstanding these developments in Si$_{3}$N$_{4}$, we note that the results presented here can be used for device design in any low-loss material with a similar index of refraction, such as diamond~\cite{Mouradian:17}.


\section{Photonic crystal nanocavity design}

\begin{figure}[!t]
\centering
\includegraphics[width=12cm]{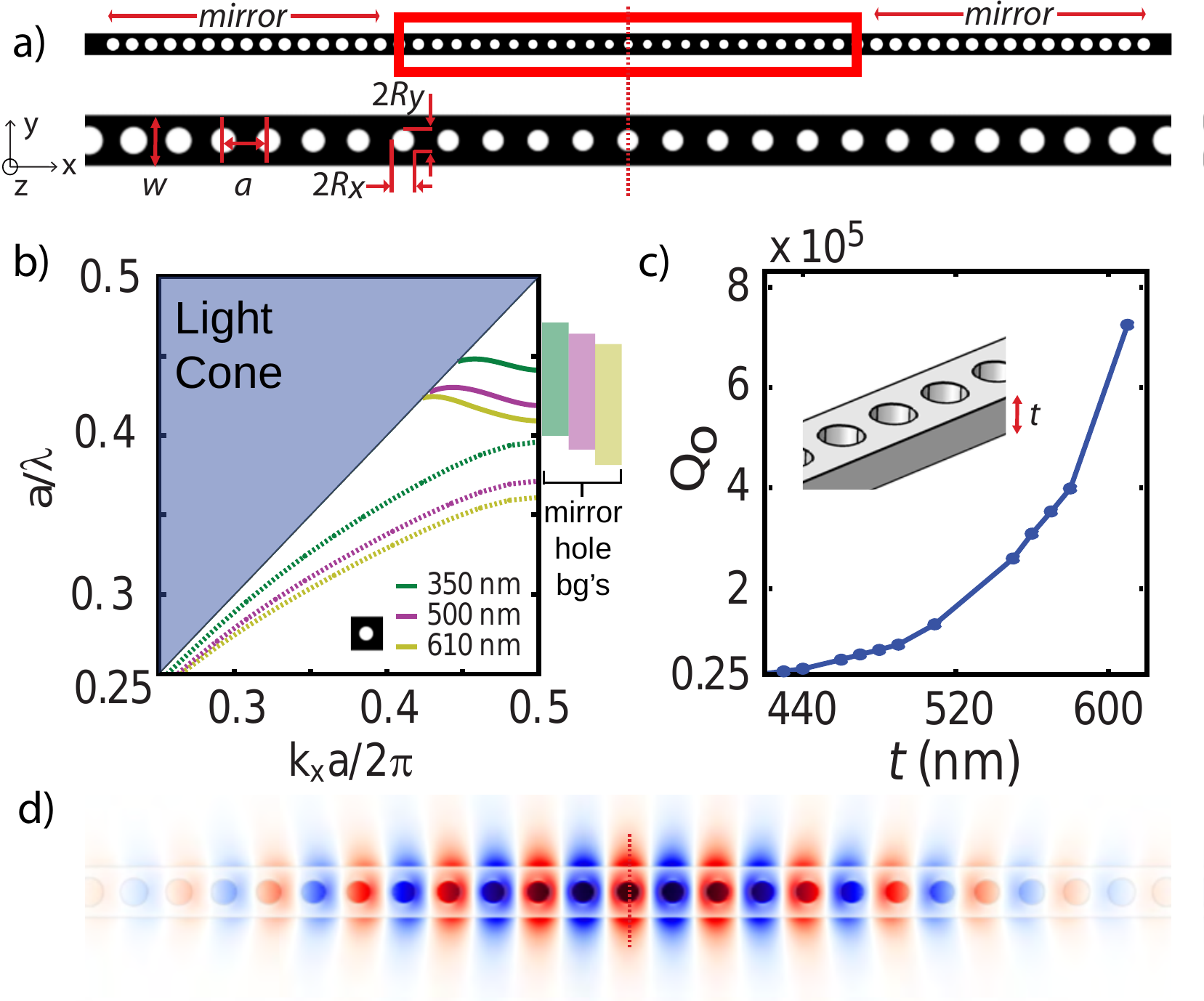}
\caption{(a) Schematic of the Si$_{3}$N$_{4}$ photonic crystal dielectric patterning with the cavity section shown enlarged and dotted line denoting the cavity centre. Dimensions: width $w$, lattice constant $a$, semi-minor and -major axes ($R_x,R_y$). (b) The TE-polarization bandstructure for the centre hole unit cell of the device in (a) for varying nanobeam thicknesses.  The air (dielectric) modes are denoted with solid (dotted) lines, and the bandgaps (bg's) of the mirror holes for the respective coloured bands are shown to the right of the plot area.  (c) The quality factor $Q_o$ of the PCC's symmetric fundamental mode determined from 3D FDTD simulations for thickness $t$. (d) The 3D FDTD electric field $E_y$ mode profile for a 500~nm thick device.}
\label{figureOne}
\end{figure}

We begin by discussing the design of the Si$_{3}$N$_{4}$ air-mode PCC shown in Fig.\ \ref{figureOne}(a). The hole dimensions and tapering were calculated with the deterministic design recipe in Refs.\ \cite{Quan10, Quan11}, as discussed below. We set the nanobeam width~$w$~=~760~nm and lattice constant~$a$~=~670~nm such that the target PCC mode wavelength~$\lambda_o$~$\sim$~1550~nm~($\omega_o~\sim~200$~THz, $a/\lambda_o~\sim$~0.432) can be achieved for the range of hole sizes considered here. Initially the nanobeam thickness was set to $t = 350~\text{nm}$, then varied to study its effect on the PCC modes.  As part of this design recipe,  the band structure and band-edge frequencies of unit cells of the waveguide for varying  hole shape were calculated using a 3D frequency-domain eigensolver~\cite{Johnson:01} with high spatial resolution.  From these bandstructure calculations, we first find the central cavity hole dimensions that brings the unit-cell air-mode band-edge frequency close to the target $\omega_o$. We then find the outer mirror hole dimensions to maximize the mirror strength $\gamma$:
\begin{equation}
\gamma=\left( \frac{(\omega_2-\omega_1)^2}{(\omega_2+\omega_1)^2}-\frac{(\omega_\textrm{o}-\omega_{mid})^2}{(\omega_{mid})^2} ) \right) ^{1/2}, 
\end{equation}
experienced by the mode at $\omega_\textrm{o}$ that is within the mirror hole bandgap, defined by the dielectric and air band-edge frequencies $\omega_1$ and $\omega_2$, respectively, of the mirror hole unit cell~\cite{Hryciw:13}. Here $\omega_{mid} = (\omega_2+\omega_1)/2$ is the mid bandgap frequency, and $\gamma$ is proportional to the inverse attenuation length.

\begin{figure}[!b]
\centering
\includegraphics[width=12cm]{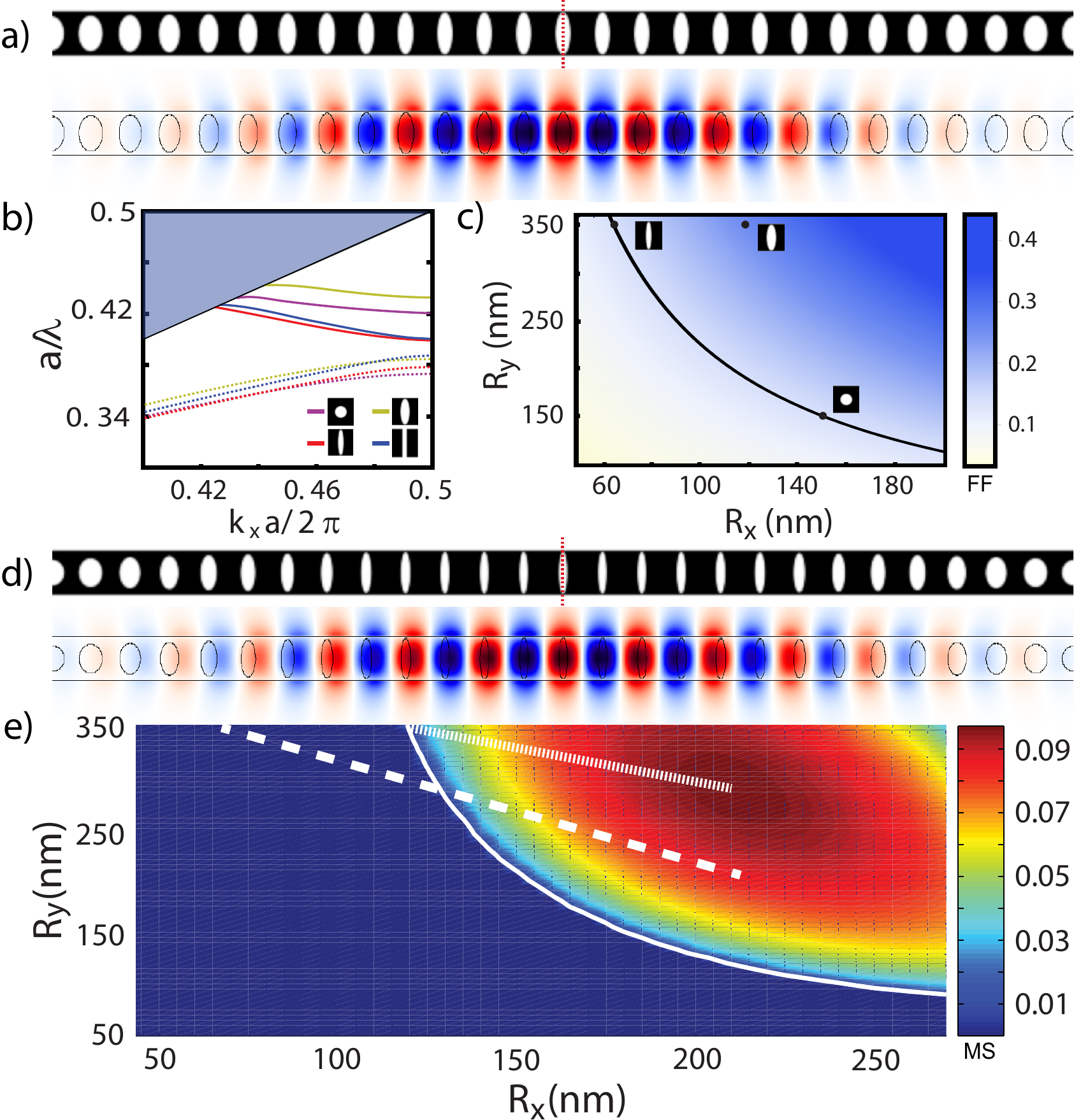}
\caption{(a) The dielectric pattern and $E_y$ field profile of the elliptical hole air-mode PCC following the standard deterministic design procedure.   (b) The band structures of a unit cell containing the three centre hole shapes and gap discussed: 150~nm round~(purple), standard design ellipse~(yellow), highly elliptical~(red), 112~nm~gap~(blue).  (c) Fill factors (FF) of ellipses with dimensions ($R_x,R_y$).  The black line denotes elliptical hole dimensions with a fill factor similar to the round centre hole.  Fill factors of the three centre hole shapes are noted, with the gap FF matching that of the highly elliptical hole.  (d) Schematic of the PCC with a highly elliptical centre hole and hole axes quadratically tapering to round mirror holes, and the resultant $E_y$ field profile. (e) Mirror strength (MS) trajectories of the hole tapering of the devices in (a) (dotted) and (d) (dashed).  The hole dimensions that have a mirror strength at the target frequency~$\omega_o$ are shown with the solid white line.}
\label{figureTwo}
\end{figure}

Initially we restricted ourselves to circular holes. A cavity centre hole with radius 150~nm was found to have an air-mode band-edge frequency $\omega_2 \sim \omega_o$. To form an air-mode nanocavity, the surrounding hole radii were tapered quadratically over $N_c$~=~12 holes symmetrically on either side of the nanobeam to the first mirror holes with radius 215~nm. $N_m$~=~15 mirror holes were then added to both sides.  The bandstructure for the centre hole unit cell plotted in Fig.~\ref{figureOne}(b) shows the fundamental (TE-like,~$y$-odd parity,~$z$-even parity)  air and dielectric mode frequencies in normalized units ($\omega a/2\pi c~=~a/\lambda$) as a function of $k_x$ for varying $t$.  From the bandstructure, we see that the air-mode band-edge frequency for the $t=350~\text{nm}$ device is relatively close to the light-line. When we compute the fundamental optical mode of the full $t=350~\text{nm}$ PCC using a 3D finite-difference time-domain (FDTD) simulation~\cite{ref:oskooi2010mff} we find a $Q_o \sim 10^4$.  A similar $Q_o$ is observed if we keep the cavity hole radii constant and quadratically decrease the lattice constant from the centre of the cavity to the mirrors~\cite{Grutter:15IEEE}.  

To increase $Q_o$ we then studied the effect of changing $t$. In nanobeam PCC's, light is index-guided in the $z$ direction, and typically $t \le \lambda / 2n$ is chosen so that the underlying waveguide remains single-mode over the frequency range of interest~\cite{Zhang:09}.  Figure~\ref{figureOne}(b) shows that increasing the thickness pushes the band structure lower in frequency, away from the light-line: as $t$ is increased to 610~nm, the air-mode band-edge falls from a normalized frequency of 0.441 to 0.409. This corresponds to increasing $n_{eff}$, which decreases $\omega_{o}$ and reduces coupling to radiation modes, and is reflected by $Q_o$ increasing towards $10^6$ as shown in Fig.~\ref{figureOne}(c).  A simulated optical mode electric field profile~($E_y$) for a  $t$~=~500~nm device is shown in Fig.~\ref{figureOne}(d), whose corresponding $Q_o$ is $\sim 10^5$. In fabricated Si$_{3}$N$_{4}$~PCC devices, $t$ generally ranges from 200-450~nm~\cite{Khan:11,Grutter:15IEEE}. In consideration of the single-mode condition and with a desire for a high-$Q_o$, we will use a $t \sim 500~\text{nm}$ for the remainder of this analysis.


To design a nanobeam PCC with elliptical holes that will form the basis of the optomechanical devices discussed below, we relax the constraint of requiring holes to be circular, leave the other nanobeam dimensions unchanged, and again apply the deterministic design procedure~\cite{Hryciw:13}. There are a continuum of hole shapes with varying ellipticity that maintain the same $\omega_o$ as the round hole device, as discussed below. We begin by choosing the centre hole with maximum ellipticity as constrained by the waveguide width. The resulting device is shown in Fig.\ \ref{figureTwo}(a).  The cavity centre hole with  semi-minor and semi-major axis dimensions ($R_{x_\textrm{c}}$,~$R_{y_\textrm{c}}$)~=~(119.4,~350)~nm has $a/\lambda~=~0.432$, and Fig.\ \ref{figureTwo}(b) shows how the bandstructure of this unit cell compares with that of the 150~nm radius round centre hole unit cell. If we quadratically taper to a mirror hole ($R_{x_\textrm{m}}$,~$R_{y_\textrm{m}}$)~=~(210,~285)~nm with maximum of the mirror strength following:
\begin{equation}
R_{x_j,y_j}= R_{x_\textrm{c},y_\textrm{c}}+ (j/N_\textrm{c})^2(R_{x_\textrm{m},y_\textrm{m}}-R_{x_\textrm{c},y_\textrm{c}})
\end{equation}
for integer $j\in[-N_\textrm{c},N_\textrm{c}]$, we find a simulated 3D~$Q_o$~=~$4.4~\times~10^4$.  

For the purpose of comparison, we also simulated a device with a highly elliptical centre hole with the same fill factor as the round centre hole of the device in Fig.\ \ref{figureOne}(a).  In Fig.\ \ref{figureTwo}(c), the fill factors of ellipses of dimensions ($R_x,R_y$) are calculated; the black line shows hole dimensions with the same fill factor as the round centre hole.  The bandstructure for the hole with dimensions ($R_{x_c},R_{y_c}$)~=~(64.3,~350)~nm is shown in Fig.~\ref{figureTwo}(b)~(red line) to have much lower bandedge frequencies; substituting this highly elliptical hole for the round cavity centre hole and tapering quadratically to the 215~nm round mirror holes results in the device shown in Fig.\ \ref{figureTwo}(d) with a 3D simulated $Q_o$~=~$5.3~\times~10^5$. This is more than an order of magnitude greater than the $Q_o$ of the PCC designed following the standard deterministic recipe based on bandedge matching. This can be explained by this design's lower mode frequency $a/\lambda=0.40$ that is further from the air light-line and as a result has less vertical radiation loss. For comparison, the trajectories of the hole mirror strengths of both the band-edge matching design~(dotted~line) and the highly elliptical to round hole fill-factor design~(dashed~line) are shown in Fig.\ \ref{figureTwo}(e). Also note the solid white equi-frequency line at the target frequency in this Figure.
We expect that numerical optimization in future work, e.g.\ using a genetic algorithm \cite{goh2007genetic}, would allow $Q_o$ to be increased further.

\section{Creating split-beam and paddle nanocavity optomechanical devices}

\begin{figure}[!t]
\centering
\includegraphics[width=11.5cm]{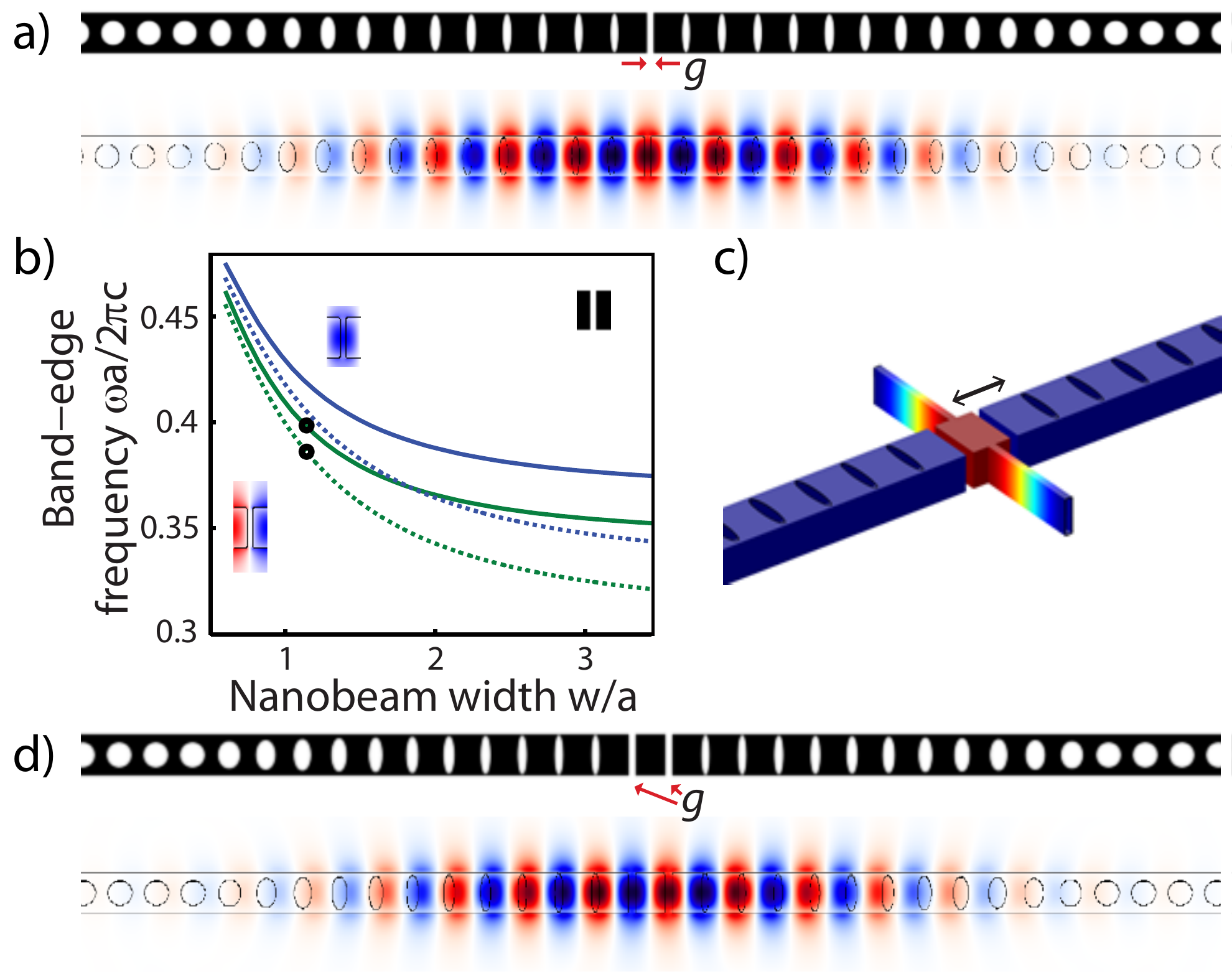}
\caption{(a) Schematic of the split-beam cavity with gap width $g$ and corresponding $E_y$ field profile.  (b)  Gap unit cell band edge frequencies (green) and $E_y$ field profiles for the optical modes at the points marked (and in blue at the same $w/t$ ratio as in Ref.~\cite{Hryciw:13}) showing that the modes for the gap for each device do not cross.  (c,d) Design and field profile of the paddle nanocavity with the axial motion of the paddle indicated (supports not shown in (d)).}
\label{figureThree}
\end{figure}

We now show how the highly elliptical hole device from Fig.~\ref{figureTwo}(d) can be used as the basis for the split-beam cavity shown in Fig.\ \ref{figureThree}(a).  A split-beam cavity is formed by making a clear cut through the centre of the nanobeam, creating nanocantilevers whose mechanical modes are similar to those used in many sensing and metrology applications. Placing the cut in the centre of PCC ensures that the nanocantilever motion interacts strongly with the optical cavity mode. In order to maintain high-$Q_o$, tapering to highly elliptical holes is beneficial, as they better match the rectangular shape of the gap~\cite{Hryciw:13}. When designing SBC's, it is critical to understand the optical mode structure to know what gap width $g$ to replace the centre hole with.  With SBC's in Si, it was previously found that the dielectric mode of the gap had to be matched to the air mode of the centre hole due to a crossing of the waveguide band edges~\cite{Hryciw:13}.  As shown in Fig.~\ref{figureThree}(b), due to the lower refractive index of the Si$_{3}$N$_{4}$, the dielectric and air mode bandedge frequencies of the gap unit cell as a function of nanobeam width in Si$_{3}$N$_{4}$ do not cross, even with the same $w/t$ ratio as in Ref.~\cite{Hryciw:13}.  Knowing this, we then substitute a gap with $g$~=~110~nm that has the same fill factor as the cavity centre hole that it replaces (given from Fig.~\ref{figureTwo}(c)). After local optimization of $g$, we find that for $g$~=~112~nm the SBC supports an air-mode with $Q_o$~=~$2.2~\times~10^5$.  The bandstructure of a unit cell containing this gap is shown in Fig.~\ref{figureTwo}~(blue line) to overlap very closely with that of the highly elliptical hole.

\begin{figure}[!t]
\centering
\includegraphics[width=11cm]{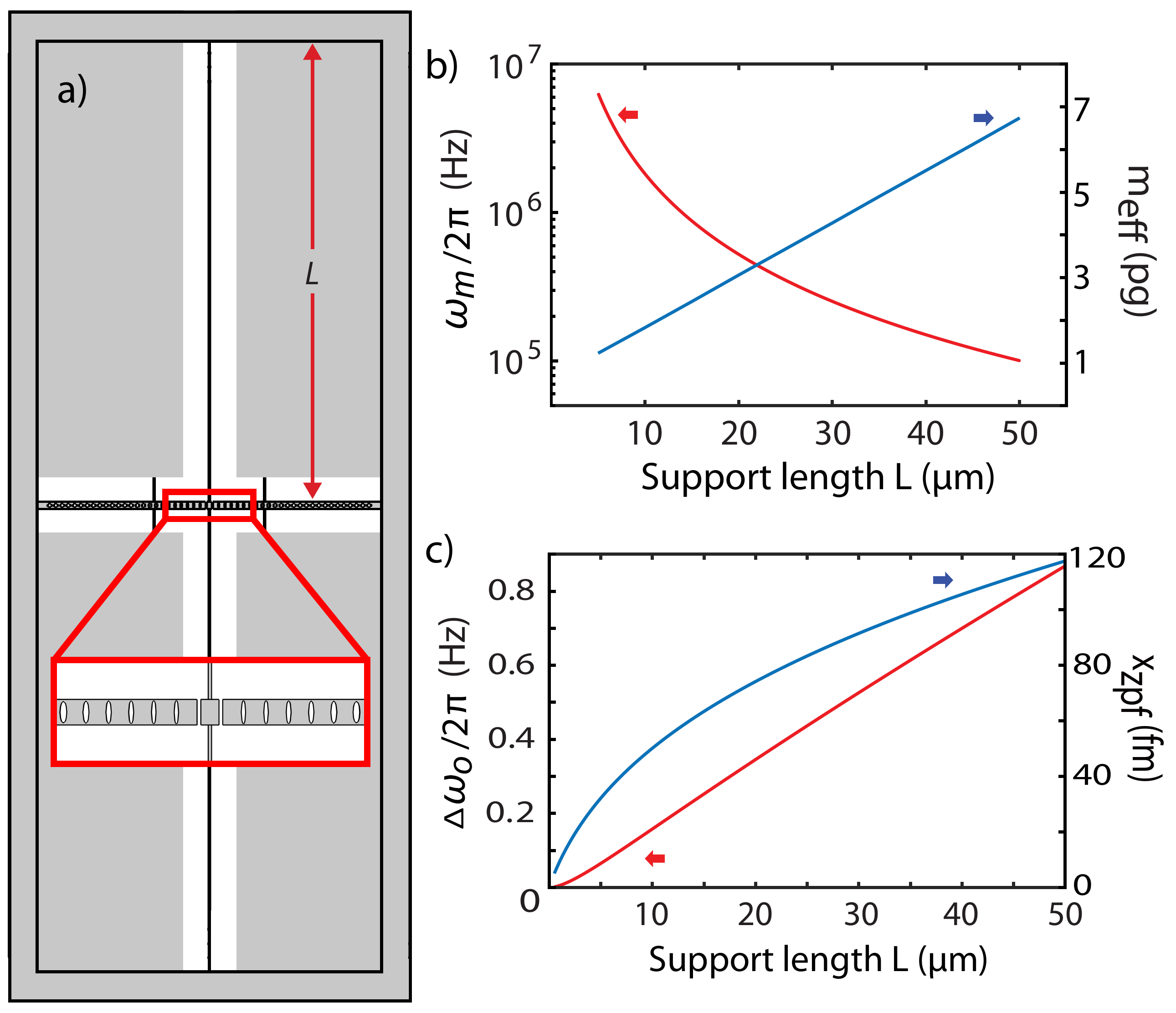}
\caption{(a)  Schematic of the paddle nanocavity with $L$~=~50~$\mu$m supports.  (b)  Mechanical frequency~$\omega_m$~(red) and effective mass~$m_{eff}$~(blue) and (c) the single-photon to two-phonon coupling rate $\Delta\omega_{o}$~(red) given from the zero-point fluctuations $x_{zpf}$~(blue).}
\label{figureFour}
\end{figure}

A similar process can be followed to create a Si$_3$N$_4$ paddle nanocavity such as the device shown in Figs.\ \ref{figureThree}(c,d). In this device a Si$_3$N$_4$ paddle is inserted between the split-beam cavity nanocantilevers, with the same gap $g = 112\text{nm}$ between each nanocantilever and the paddle. The paddle length is chosen following the procedure used to design Si paddle nanocavities in Ref.\ \cite{Healey:15}. When simulated with a paddle length of $\sim$538~nm and without any supports, as shown in Fig.~\ref{figureThree}(d), a $Q_o$~=~$1.1~\times~10^5$ is found. In a realistic device, the paddle is suspended from the unpatterned device layer by thin supports as shown in Fig.\ \ref{figureThree}(c). Adding 100~nm wide supports connected to the center of the paddle reduces $Q_o$~to~$1.7~\times~10^4$. 

The support and paddle dimensions determine the properties of the paddle's mechanical resonances, which can interact optomechanically with the optical modes of the device. Here we are interested in the axial sliding mechanical resonance shown in Fig.~\ref{figureThree}(c) that mimics membrane-in-the-middle devices~\cite{Thompson2008, sankey2010strong, mason2018continuous,rossi2018observing,rossi2018measurement}.  For the sliding resonance, the paddle nanocavity has a vanishing linear optomechanical coupling owing to the odd symmetry of the mechanical mode and the even symmetry of the optical mode intensity \cite{Kaviani:15}. However, it has a nonlinear quadratic optomechanical coupling that is of particular interest for quantum optomechanics experiments \cite{Thompson2008, sankey2010strong, paraiso2015position}. This coupling strength is  determined by the change in local dielectric constant $\Delta\epsilon(\mathbf{r};x)$ as a function of position $\mathbf{r}$ of the paddle's normal mechanical mode displacement amplitude $x$, overlapped and cross-coupling the electric fields $E_{\omega}$ of the nanocavity optical mode spectrum at frequencies $\omega$~\cite{Kaviani:15}.  The quadratic optomechanical coupling coefficient:
\begin{equation}
g^{(2)}=\frac{\omega}{2} \frac{|\langle E_{\omega}|\frac{\delta \epsilon}{\delta x}|E_{\omega}\rangle|^2}{|\langle E_{\omega}| \epsilon |E_{\omega}\rangle|^2} -\sum\limits_{\omega^{\prime}\neq\omega}\left(\frac{\omega^{3}}{\omega^{\prime 2}-\omega^{2}}\right) \frac{|\langle E_{\omega^{\prime}}|\frac{\delta \epsilon}{\delta x} |E_{\omega}\rangle|^2}{\langle E_{\omega^{\prime}}| \epsilon |E_{\omega^{\prime}}\rangle \langle E_{\omega}| \epsilon |E_{\omega}\rangle}
\label{eq:second_order}
\end{equation}
describes the strength of the nonlinear photon-phonon interactions in the system~\cite{Johnson02,Rodriguez:11,Eichenfield:09OE}.  As discussed above, the first self-term  describing the nonlinear contribution from the linear optomechanical coupling vanishes~\cite{Kaviani:15}.  However, in practice there is some amount of linear coupling~\cite{Healey:15} generated by imperfections in the fabricated device. Linear coupling can also be introduced with an external waveguide used to couple light into and out of the cavity~\cite{Hryciw:15}.  Following Ref.\ \cite{Kaviani:15}, a $g^{(2)}/2\pi\approx$~10~MHz/nm$^{2}$ was calculated by inputting the mechanical sliding resonance profile and the optical field profiles of the fundamental ($E_\omega$) and higher order ($E_{\omega'}$) optical modes of the paddle nanocavity into Eq.~\ref{eq:second_order}. These calculations where performed using finite element method simulations (COMSOL), and  contributions from higher order modes over a spectral range of tens of terahertz were considered. Including this extended higher order mode spectrum was aided by automating the calculations that were performed manually in Ref.~\cite{Kaviani:15}.


\begin{figure}[!t]
\centering
\includegraphics[width=11.5cm]{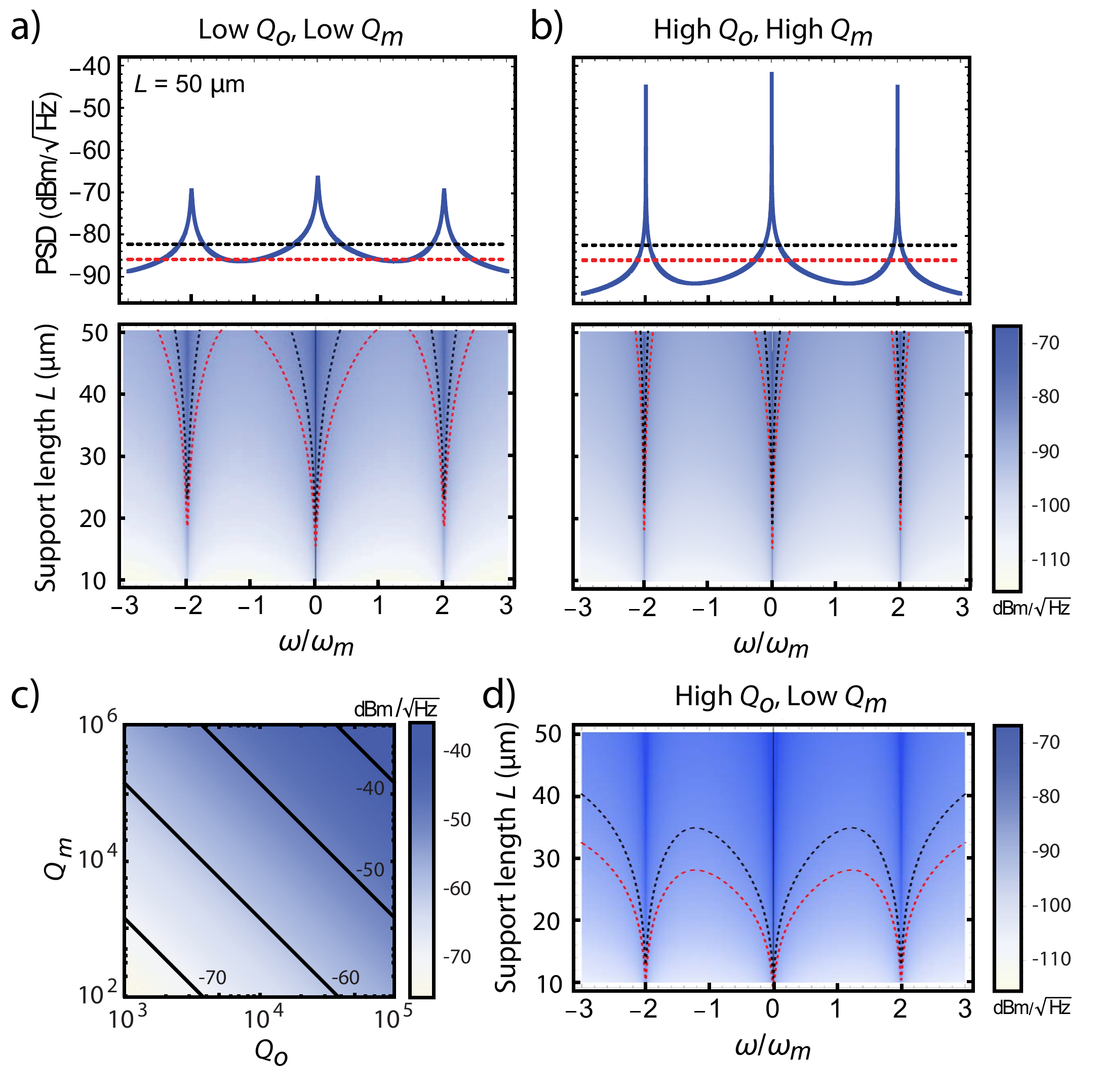}
\caption{PSD of the mechanical resonance for (a)~low $Q_o$~=~$4.0~\times~10^3$, $Q_m~=~10^2$ and (b)~high $Q_o$~=~$4.0~\times~10^4$, $Q_m~=~10^5$ ((a,b) top plots are with $L \sim 50~\mu$m long supports).  The detector noise (lower, red dashed line) and shot noise (upper, black dashed line) are also shown.  (c)~The PSD at $L$~=~50~$\mu$m and $\omega/\omega_m$~=~2.  (d)~PSD for high $Q_o$, low~$Q_m$ as given in (a,b).}
\label{figureFive}
\end{figure}

Although $g^{(2)}$ is reduced in Si$_3$N$_4$ paddle nanocavities in comparison to their Si implementation, due in part to the lower refractive index of Si$_3$N$_4$, applications in quantum optomechanics can benefit from the exceptionally high aspect-ratio mechanical structures realizable in Si$_3$N$_4$. To illustrate this potential, we now investigate the effect of increasing the paddle's sliding mode per-phonon displacement amplitude by increasing the support length~$L$ to reduce the mode's spring constant.  Figure~\ref{figureFour}(a) shows a schematic of the device with a 50~$\mu$m~$\times$~100~nm support cross section.  The mechanical frequency~$\omega_m$  and effective mass~$m_{eff}$ of the sliding mode (computed with COMSOL) as a function of support length are shown in Fig.~\ref{figureFour}(b). For $L$~ranging~5-50~$\mu$m,  $\omega_m/2\pi$ varies from $\sim10^7-10^5$~Hz, and has a $L^{-1.9}$ dependence slightly shifted from $L^{-2}$ as expected from Euler-Bernoulli nanobeam theory~\cite{Cleland:02} due to the mass distribution of the paddle.  The corresponding $m_{eff}$ varies from $\sim1-7$~pg.  Shown in Fig.~\ref{figureFour}(c), the zero-point fluctuation amplitude $x_{zpf}~=~\sqrt{\hbar/2m_{eff}\omega_{m}}$ varies from 20-120~fm (where $\hbar$ is the reduced Planck's constant), and the corresponding single-photon to two-phonon coupling rate $\Delta\omega_{o}\equiv|g^{(2)}x^{2}_{zpf}|$  varies from 2$\pi~\times$~(0.25-0.9)~Hz.  These plots illustrate that increasing $L$ results in a larger per-phonon nonlinear optomechanical coupling due to the increase in $x_{zpf}$.

Given these simulated parameters we can make predictions regarding the suitability of Si$_3$N$_4$ paddle nanocavities for nonlinear optomechanical detection of motion driven by thermal and quantum effects. If photons are coupled into and out of the cavity with an external waveguide (e.g. a fiber taper or end-fire coupled waveguide) and measured with a photodetector, the optical response of the paddle nanocavity transducing the nonlinear mechanical displacement can be predicted~\cite{Kaviani:15}. For input power~$P_i$, the photodetected optical power spectral density~(PSD)~$S_{P}^{(2)}(\omega) = \frac{1}{4}P_i^2 G_2^{2} S_{x^{2}}(\omega)$, where $G_2$ is the quadratic coefficient of transduction of the fluctuating waveguide output, and $S_{x^{2}}$ is the PSD of the $x^2$ mechanical motion, given in Ref.~\cite{Kaviani:15} for the case of a thermally driven mechanical resonator.  Here $G_2$ describes the cavity optical response measured by the photodetector, and includes the influence of the photodiode quantum efficiency, $Q_o$, $g^{(2)}$, and waveguide transmission losses. Both $G_2$ and $S_{x^{2}}$ assume there is no backaction and the device is operating in the sideband-unresolved regime~\cite{ref:aspelmeyer2014co}. 

Figures~\ref{figureFive}(a,b) show the thermally driven $S_{P}^{(2)}(\omega)$ of the sliding resonance at an operating temperature of 300~K, and indicate that it can be observed above technical noise.  The signal strength varies with $L$ and the mechanical detuning~$\omega/\omega_m$; traces in Fig.'s~\ref{figureFive}(a,b) correspond to signal for $L$~=~50~$\mu$m in their respective spectrographs $S_{P}^{(2)}(\omega,L)$, which are also shown. These spectra assume reasonable experimental parameters:  $P_i$~=~100~$\mu$W chosen so as to not saturate the detector, the input laser detuning from the nanocavity resonance  set to maximize $G_2$, and either relatively low $Q_o$~=~$4.0~\times~10^3$ and $Q_m$~=~$10^2$~(Fig.'s~\ref{figureFive}(a)) or attainably high $Q_o$~=~$4.0~\times~10^4$ and $Q_m$~=~$10^5$~(Fig.'s~\ref{figureFive}(b)).  The power spectrum is strongest at $\omega = 0$ and $\omega/\omega_m$~=~$\pm$~2, with the latter being of interest in practical experiments so as to minimize the influence of $1/\omega$ technical noise not considered here. The signal generally increases with $Q_o$ and $Q_m$, and the FWHM (full-width, half-maximum) of the peaks narrows with increasing $Q_m$.  As shown in Fig.\ \ref{figureFive}(b), for the high $Q_o$ and $Q_m$ values, the nonlinear signal can exceed the electronic noise (red dashed line) of a Newport~1811~photoreceiver (NEP~=~2.5~pW/Hz) and the optical shot noise (black dashed line) by over 30~dBm/$\sqrt{\text{Hz}}$. 

The importance of increasing $L$ can be seen from the spectrographs in Figs.\  \ref{figureFive}(a,b), whose contours indicate that the frequency range where the signal exceeds the noise sources becomes larger as $L$ increases. This is a result of the increase in thermal occupancy (due to decreasing $\omega_m$) and increase in $g^{(2)}$ of the device with increasing $L$. The impact of reducing mechanical and optical dissipation is illustrated by Fig.\ \ref{figureFive}(c), which shows how the signal strength at $L$~=~50~$\mu$m and $\omega/\omega_m$~=~2 depends on $Q_o$ ranging over $10^3-10^5$ and $Q_m$ over $10^2-10^6$.  At the maximum of these values, the signal strength reaches $\sim-35$~dBm/$\sqrt{\text{Hz}}$, approximately 50~dBm/$\sqrt{\text{Hz}}$ over the predicted system noise levels.  Figure~\ref{figureFive}(d) shows that increasing $Q_o$ while maintaining low $Q_m$ raises the signal level but maintains a broad bandwidth, making the signal accessible at all mechanical frequencies between $\omega/\omega_m$~=~$\pm$3 for support lengths~\textgreater~40~$\mu$m.

\begin{figure}[!t]
\centering
\includegraphics[width=12.5cm]{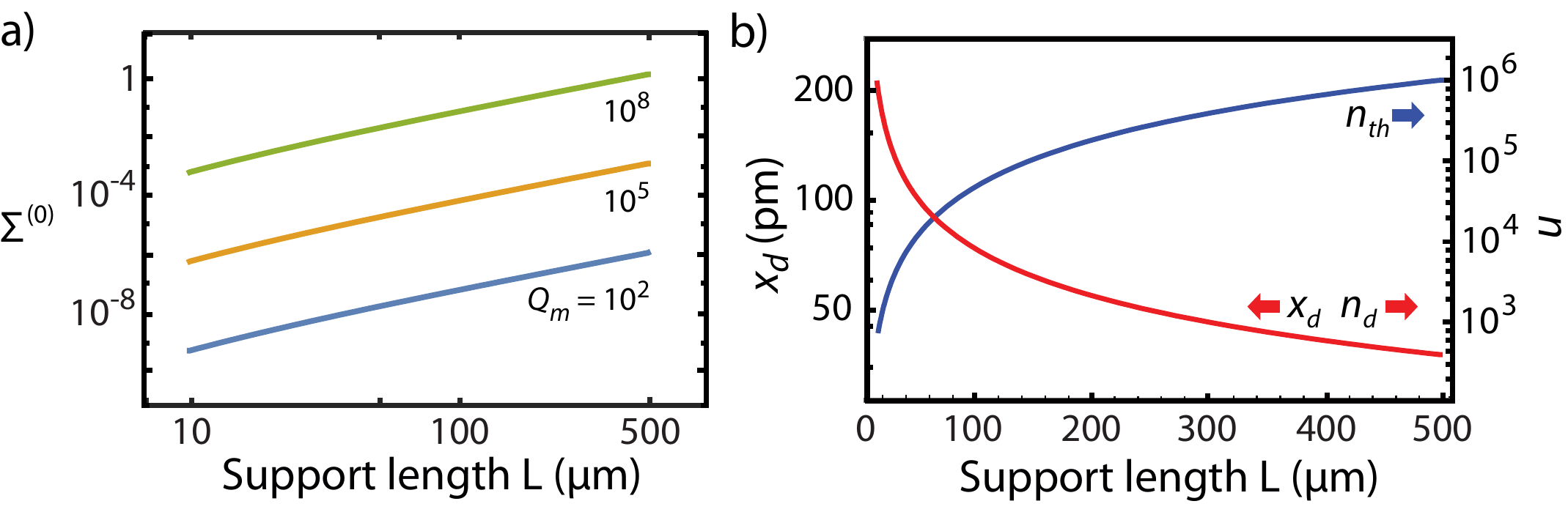}
\caption{(a) Predicted phononic Fock state measurement signal-to-noise ratio~$\Sigma^{(0)}$ for $Q_m$~=~$10^8$ (top, green), $Q_m$~=~$10^5$ (middle, orange), and $Q_m$~=~$10^2$ (bottom, blue) at ultrahigh~$Q_o$~=~$10^6$.  (b) The number of phonons $n_{th}$ in the device at $T_b$~=~10~mK~(blue) and the drive amplitude $x_d$~(red, and in units of phonon number $n_d$) to achieve a signal-to-noise ratio~$S$~\textgreater~1 when measuring phonon shot noise.}
\label{figureSix}
\end{figure}

Finally, we discuss how the Si$_{3}$N$_{4}$ paddle nanocavity is expected to perform for phononic quantum non-demolition (QND) measurements~\cite{Kaviani:15}.  Although the $g^{(2)}$ for this device is almost two orders of magnitude lower than that of similar Si devices, the high internal stress of Si$_{3}$N$_{4}$ enables longer supports that in practice lower $\omega_m$. Stress engineering can be used to increase $Q_m$  and has resulted in record $Q_m \omega_m$ products in recently reported work \cite{ghadimi2018elastic, tsaturyan2017ultracoherent}. This figure of merit is a measure of the ability of a mechanical resonator to maintain coherence in the presence of a thermal bath.  Considering measurement shot noise and the expected change in cavity transmission due to a change in phonon number, as in  Ref.\ \cite{Thompson2008}, it is predicted that the signal-to-noise ratio  $\Sigma^{(0)}$~=~$\tau_{tot}^{(0)}\Delta\omega_{o}^{2}/S_{w_{o}}$ for a  quantum jump measurement from the motional ground state to a single phonon Fock state could exceed unity for sufficiently high $Q_m$ and $L$, as shown in Fig.~\ref{figureSix}(a).  The $\Sigma^{(0)}$ for $Q_m$~=~$10^2$,~$10^5$ and $10^8$~(blue, orange and green curves respectively) is presented, assuming the device and its surroundings are cryogenically pre-cooled to a bath temperature $T_{b}~=~10$~mK, and that $Q_o$~=~$10^6$. Here $\tau_{tot}^{(0)}~=~\hbar Q_m~/~k_{B}T_{b}$ (for Boltzmann's constant $k_B$) is the thermal lifetime quantifying the rate of decoherence of the ground-state-cooled nanomechanical resonator due to bath phonons, and $S_{w_{o}}~=~\hbar\omega_{o}\kappa^{2}/16P_{i}$ is the shot-noise-limited sensitivity of an ideal Pound-Drever-Hall detector for cavity loss rate~$\kappa$~(cooling the device to the ground state will be further discussed below).  With $Q_m$~=~$10^8$ and $L$~=~500~$\mu$m, $\Sigma^{(0)}$~\textgreater~1, however this can also be achieved with lower $Q_m$ and longer $L$. We note that calculating the signal-to-noise ratio as in Ref.~\cite{paraiso2015position} gives similar results.

For measurements of phonon shot noise (the granularity of large phonon number coherent states)~\cite{ClerkPRL2010} with signal-to-noise ratio~$S=8\bar{n}_{d}n_{th}\Sigma^{(0)}$~\textgreater~1, the required drive amplitude~$x_d$ and its correspondence in units of phonon number~$\bar{n}_d$, assuming a thermal phonon number~$n_{th}=1/4$, is shown in Fig.~\ref{figureSix}(b).  At $L$~=~500~$\mu$m, the number of required coherent drive phonons is \textless~400, a relatively small value that is more than three orders of magnitude less than what is required to drive the Si device previously discussed~\cite{Kaviani:15}. This leaves significant available headroom in drive strength to compensate for non-optimal devices parameters and reduced $\Sigma^{(0)}$.

However, the above figures of merit do not consider the effect of parasitic linear optomechanical coupling between the optical cavity modes that is discussed in detail in Refs.\ \cite{miao2009standard, ludwig2012enhanced, paraiso2015position}. This coupling, which originates from the non-zero cross-coupling terms in Eq.\ \eqref{eq:second_order}, is a form of decoherence that imposes a strong coupling requirement for QND measurements \cite{miao2009standard}. Applying the theory from Ref.\ \cite{miao2009standard} to our system and assuming that the nonlinear coupling is dominated by a single cross-coupling term, this condition roughly requires $\kappa_1\kappa_2/|\Delta\omega_o (\omega'~-~\omega)|~<~1$, where $(\omega'~-~\omega)$ is the detuning between the cavity modes whose cross-coupling is the dominant contributor to $g^{(2)}$ in Eq.\ \eqref{eq:second_order}, and $\kappa_i = \omega_i/Q_i$ are the optical loss rates of the modes. For our system, $|\omega' - \omega|/2\pi \sim 10$ THz, $\kappa_{1,2}/2\pi \sim 1$ GHz ($Q_o~\sim~10^5)$, and $g^{(2)}/2\pi \sim 1$ Hz, indicating that the device is four to five orders of magnitude away from satisfying this condition, and that increasing $Q_o$ as well as $g^{(2)}$ and/or $|\omega - \omega'|$ is required before QND measurements will be possible. Increasing $Q_o$ to $10^7$, as achieved in silicon photonic crystals \cite{sekoguchi2014photonic}, is one such path towards this goal. In parallel, further increasing $g^{(2)}$ while maintaining a large $|\omega~-~\omega'|$ would contribute to reaching this regime.

Although the longer supports make the quantized phononic energy fluctuations of the oscillator observable, there will be increased thermal decoherence associated with the lower~$\omega_m$.  There is also a trade-off in the increase in thermal phonon occupation number at achievable cryogenic operation temperature, requiring development of the ability to further cool the mechanical resonance from $T_b$ to an effective temperature corresponding to the quantum ground state.  The number of phonons~$n_{th}$ remaining in the device at $T_{b}$ and $\omega_m$ given by $L$ (according to Bose-Einstein statistics) is shown in Fig.~\ref{figureSix}(b)~(blue). Removing these phonons in order to demonstrate the quantum effects described above would require sideband-unresolved nonlinear optomechanical cooling~\cite{Clark2017} to extract mechanical energy by scattering incident photons to higher energies~(anti-Stokes scattering), as recently achieved in a low-frequency optomechanical system~\cite{Clark2017}.  By interacting squeezed light with the optomechanical cavity, one can coherently null the Stokes processes which prevent cooling, and the mechanical system can in principle be cooled arbitrarily close to the motional quantum-mechanical ground state~\cite{Clark2017}.  Techniques to increase the $Q_m$, such as the utilization of phononic crystals~\cite{Chan2011} and nanostring strain engineering \cite{ghadimi2018elastic,ghadimi2017radiation,fedorov2018evidence}, could also help reach this goal.



\section{Conclusion}

Designing a silicon nitride air-mode PCC is a delicate task due to the sensitivity of the device to optical loss through radiating modes.  High optical quality factor devices can be realized by reducing the air-mode bandedge frequencies through increasing the device thickness.  These devices can form the basis for optomechanical devices that can be used for both fundamental and practical research and applications. 


\section*{Acknowledgments}
The authors are grateful to Roohollah Ghobadi for helpful discussions.  This work is supported by Alberta Innovates~(AI), the Natural Science and Engineering Research Council of Canada~(NSERC), and the Canada Foundation for Innovation~(CFI).

\section*{Disclosures}

The authors declare that there are no conflicts of interest related to this article.

\end{document}